\pdfoutput=1

\RequirePackage{fix-cm}

\documentclass[%
floatfix,
showkeys,
nofootinbib, %
superscriptaddress, %
]{revtex4-1}

\usepackage{cmap}

\usepackage{ucs}
\usepackage[utf8x]{inputenc}
\usepackage[T1,T2A]{fontenc}
\usepackage[german,russian,english]{babel}

\usepackage[sort&compress]{natbib}
\usepackage{amsmath}
\usepackage{amssymb}

\usepackage{mathtools}
\mathtoolsset{
showonlyrefs,
mathic = true
}

\allowdisplaybreaks

\usepackage{hyperref}
\hypersetup{backref,
 colorlinks=false}
\hypersetup{pdfborder=0 0 0}

\usepackage{microtype}
\UseMicrotypeSet[protrusion]{alltext}

\usepackage{graphicx}

\usepackage[scanall]{psfrag}

\usepackage{listings}
\usepackage{listingsutf8}
\usepackage{fixltx2e}

\usepackage{nicefrac}

\makeatletter
\def\ps@pprintTitle{%
     \let\@oddhead\@empty
     \let\@evenhead\@empty
     \let\@oddfoot\@empty
     \let\@evenfoot\@oddfoot}
\makeatother

\graphicspath{{image/eikonal-sfm/en/}{image/eikonal-sfm/}{image/}}

\begin{document}

\title{Numerical analysis of eikonal equation}

\author{D. S. Kulyabov}
\email{kulyabov-ds@rudn.ru}
\affiliation{Department of Applied Probability and Informatics,\\
  Peoples' Friendship University of Russia (RUDN University),\\
  6 Miklukho-Maklaya St, Moscow, 117198, Russian Federation}
\affiliation{Laboratory of Information Technologies\\
  Joint Institute for Nuclear Research\\
  6 Joliot-Curie, Dubna, Moscow region, 141980, Russia}

\author{A. V. Korolkova}
\email{korolkova-av@rudn.ru}
\affiliation{Department of Applied Probability and Informatics,\\
  Peoples' Friendship University of Russia (RUDN University),\\
  6 Miklukho-Maklaya St, Moscow, 117198, Russian Federation}

\author{T. R. Velieva}
\email{velieva_tr@rudn.university}
\affiliation{Department of Applied Probability and Informatics,\\
  Peoples' Friendship University of Russia (RUDN University),\\
  6 Miklukho-Maklaya St, Moscow, 117198, Russian Federation}

\author{M. N. Gevorkyan}
\email{gevorkyan-mn@rudn.ru}
\affiliation{Department of Applied Probability and Informatics,\\
  Peoples' Friendship University of Russia (RUDN University),\\
  6 Miklukho-Maklaya St, Moscow, 117198, Russian Federation}

\begin{abstract}
  The Maxwell equations have a fairly simple form. However, finding
  solutions of Maxwell's equations is an extremely difficult
  task. Therefore, various simplifying approaches are often used in
  optics. One such simplifying approach is to use the approximation of
  geometric optics. The approximation of geometric optics is
  constructed with the assumption that the wavelengths are small
  (short-wavelength approximation). The basis of geometric optics is
  the eikonal equation. The eikonal equation can be obtained from the
  wave equation (Helmholtz equation). Thus, the eikonal equation
  relates the wave and geometric optics. In fact, the eikonal equation
  is a quasi-classical approximation (the Wentzel--Kramers--Brillouin
  method) of wave optics.  This paper shows the application of
  geometric methods of electrodynamics to the calculation of optical
  devices, such as  Maxwell and Luneburg lenses. The eikonal equation,
  which was transformed to the ODE system by the method of
  characteristics, is considered. The resulting system is written for
  the case of Maxwell and Luneburg lenses. 
\end{abstract}


\maketitle

\section{Introduction}

In this article, we consider the approach to transform the eikonal
equations to the ODE system. The first part of the article describes
in detail all the mathematical calculations. In the second part we
briefly describe Maxwell and Luneberg lenses, and explain the
approach to their numerical modeling, which allows to obtain ray
trajectories and wave fronts from sources of different shapes.

\section{Application of the characteristics method to the eikonal equation solution}

\subsection{The eikonal equation}

The eikonal equation can be obtained from Maxwell's equations, written
for the regions free of currents and charges, and under the condition
of a time-changing harmonic electromagnetic field in a nonconducting
isotropic
medium~\cite{born-wolf:principles_optics::en,stratton:1948::en,ll:2::en,ll:8::en}. In
general, the eikonal equation is written as a partial differential
equation of the first order:
\[
  \left\{
  \begin{aligned}
    &|\nabla u(\mathbf{r})|^{2} = n^{2}(\mathbf{r}), \quad \mathbf{r}\in\mathbb{R}^{3},\\
    & u(\mathbf{r}) = \varphi(\mathbf{r}), \quad \mathbf{x}\in\Gamma\subset\mathbb{R}^{3}.\\
  \end{aligned}
  \right.
\]
where $\mathbf{r} = (x,y,z)^{T}$ is radius-vector,
$\varphi(\mathbf{r})$ is the boundary condition, $n(\mathbf{r})$ is
the refractive index of the medium. The function $u(\mathbf{r})$ is
the real scalar function with a physical meaning of time. It is also
often called the \emph{eikonal}
function~\cite{bruns:1895,klein:1901:eikonal}.

For visualization of lens modeling results, we will consider their
projection on the $Oxy$ plane. In this case, the eikonal equation is
reduced to the following two-dimensional form:
\begin{equation}
  \label{eq:Eikonal}
  \left\{
  \begin{aligned}
 &   \left(\dfrac{\partial u(x,y)}{\partial x}\right)^{2} + \left(\dfrac{\partial u(x,y)}{\partial y}\right)^{2} = n^{2}(x,y), \quad (x,y)\in\mathbb{R}^{2},\\
 &   u(x,y) = \varphi(x,y), \quad (x,y)\in\Gamma\subset\mathbb{R}^{2}.
\end{aligned}
\right.
\end{equation}

Using the method of characteristics, the eikonal equation can be
transformed into an ODE system that can be solved by standard
numerical methods.

\subsection{Characteristics of the eikonal equation}

Let us briefly describe the method of characteristics~\cite{jeong:2007:eikonal_parallel,kimmel:1998:geodesic,zhao:2004:fsm,beliakov:1996:numerical_luneburg,gremaud:2006:fsm_eikonal,bak:2010:fast-sweeping-method} 
and the
application of this method to the eikonal equation.

The partial differential equation of the following form is considered:
\begin{equation}
  \label{eq:eq}
  a_{1}(x,y)\frac{\partial u(x,y)}{\partial x} + a_{2}(x,y)\frac{\partial u(x,y)}{\partial y} = f(x,y),
\end{equation}
where $a_{1}(x,y)$, $a_{2}(x,y)$, $u(x,y)$ and $f(x,y)$ are
sufficiently smooth functions. This equation is equivalent to the
statement that a vector field with components $a_{1}(x,y)$,
$a_{2}(x,y)$, $f(x,y)$ is tangent to the surface $z=u(x,y)$, which has
a normal vector with components $(u_{x},u_{y},-1)$. Accordingly, for
this equation one can write the system of ODE, called \emph{equations
  of characteristics}. It has the following form:
\[
  \frac{\mathrm{d} x}{\mathrm{d}t} = a_{1}(x,y),\quad
  \frac{\mathrm{d}y}{\mathrm{d}t} = a_{2}(x,y),\quad
  \frac{\mathrm{d}u(x,y)}{\mathrm{d}t} = f(x,y). 
\]
This ODE system reduces the solution of the partial differential
equation of the first order to the solution of the ODE system of the
first order.

To get the equations of characteristics for the eikonal
equation one has to perform
two steps. At the first stage, the equation should be converted to the
form~\eqref{eq:eq}, and after that the ODE system may be written
down. For the two-dimensional case the conversion of the eikonal
equation to~\eqref{eq:eq} is performed by replacing
\[
  p_{1} = \frac{\partial u}{\partial x},\quad
  p_{2} = \frac{\partial u}{\partial y}.
\]
In this case, the equation itself is converted to form:
\[
  |\mathbf{p}|^{2} = p^{2}_{1} + p^{2}_{2} = n^{2}(x,y).
\]
A number of changes should be made.
\begin{gather*}
  \frac{\partial}{\partial x}(p_1^2 + p_2^2) = 2p_1\frac{\partial
    p_1}{\partial x} + 2p_2\frac{\partial p_2}{\partial x} =
  2n\frac{\partial n}{\partial x},\\ 
  \frac{\partial}{\partial y}(p_1^2 + p_2^2) = 2p_1\frac{\partial
    p_1}{\partial y} + 2p_2\frac{\partial p_2}{\partial y} =
  2n\frac{\partial n}{\partial y}.
\end{gather*}

After that the following system of equations is obtained:
\[
  \begin{aligned}
    & p_1\dfrac{\partial p_1}{\partial x} + p_2\dfrac{\partial
      p_2}{\partial x} = n\dfrac{\partial n}{\partial x},\\
    & p_1\dfrac{\partial p_1}{\partial y} + p_2\dfrac{\partial
      p_2}{\partial y} = n\dfrac{\partial n}{\partial y},
  \end{aligned}
  \quad 
  \Longrightarrow
  \quad
  \begin{aligned}
    & \left(\mathbf{p}, \dfrac{\partial \mathbf{p}}{\partial x}\right)
    = n\dfrac{\partial n}{\partial x},\\ 
    & \left(\mathbf{p}, \dfrac{\partial \mathbf{p}}{\partial y}\right)
    = n\dfrac{\partial n}{\partial y}. 
  \end{aligned}
\]

Since
\[
  \frac{\partial p_{1}}{\partial y} = \frac{\partial^{2}
    u(x,y)}{\partial y \partial x} = \frac{\partial^{2}
    u(x,y)}{\partial x \partial y} = \frac{\partial p_{2}}{\partial
    x}, 
\]
then
\[
  \frac{\partial p_{1}}{\partial y} = \frac{\partial p_{2}}{\partial x}.
\]

Using this equality our expressions may be converted in
\[
  \begin{aligned}
   & \dfrac{\partial \mathbf{p}}{\partial x} = \left(\frac{\partial
        p_{1}}{\partial x}, \frac{\partial p_{2}}{\partial x}\right) =
    \left(\frac{\partial p_{1}}{\partial x}, \frac{\partial
        p_{1}}{\partial y}\right) = \frac{\partial p_{1}}{\partial
      \mathbf{x}} = \nabla p_{1},\\ 
    & \dfrac{\partial \mathbf{p}}{\partial y} = \left(\frac{\partial
        p_{1}}{\partial y}, \frac{\partial p_{2}}{\partial y}\right) =
    \left(\frac{\partial p_{2}}{\partial x}, \frac{\partial
        p_{2}}{\partial y}\right) = \frac{\partial p_{2}}{\partial
      \mathbf{x}} = \nabla p_{2}. 
  \end{aligned}
\]

As a result:
\[
  \begin{aligned}
    &\left(\mathbf{p}, \dfrac{\partial \mathbf{p}}{\partial x}\right) =
    n\dfrac{\partial n}{\partial x},\\ 
    &\left(\mathbf{p}, \dfrac{\partial \mathbf{p}}{\partial y}\right) =
    n\dfrac{\partial n}{\partial y},
  \end{aligned}
  \quad
  \Longrightarrow
  \quad
  \begin{aligned}
   & \left(\mathbf{p}, \nabla p_{1}\right) = n\dfrac{\partial n}{\partial x},\\
   & \left(\mathbf{p}, \nabla p_{2}\right) = n\dfrac{\partial n}{\partial y}.
  \end{aligned}
\]

Thus, the goal is achieved --- the equation~\eqref{eq:Eikonal} is
transformed in two equations of the form~\eqref{eq:eq}. 
\[
  \begin{aligned}
   & p_1\dfrac{\partial p_1}{\partial x} + p_2\dfrac{\partial
      p_1}{\partial y} = n\dfrac{\partial n}{\partial x},\\ 
   & p_1\dfrac{\partial p_2}{\partial x} + p_2\dfrac{\partial
      p_2}{\partial y} = n\dfrac{\partial n}{\partial y}, 
  \end{aligned}
  \quad
  \Longrightarrow
  \quad
  \left\{
  \begin{aligned}
    & \dfrac{p_1}{n^2}\dfrac{\partial p_1}{\partial x} +
    \dfrac{p_2}{n^2}\dfrac{\partial p_1}{\partial y} =
    \dfrac{1}{n}\dfrac{\partial n}{\partial x},\\ 
    & \dfrac{p_1}{n^2}\dfrac{\partial p_2}{\partial x} +
    \dfrac{p_2}{n^2}\dfrac{\partial p_2}{\partial y} =
    \dfrac{1}{n}\dfrac{\partial n}{\partial y}.
  \end{aligned}
  \right.
\]

The characteristics for each equation may be written down:
{\renewcommand{\arraystretch}{2.5}%
\begin{center}
  \begin{tabular}{c||c}
    $
    \dfrac{p_1}{n^2}\dfrac{\partial p_1}{\partial x} + \dfrac{p_2}{n^2}\dfrac{\partial p_1}{\partial y} = \dfrac{1}{n}\dfrac{\partial n}{\partial x}
    $&
       $
       \dfrac{p_1}{n^2}\dfrac{\partial p_2}{\partial x} + \dfrac{p_2}{n^2}\dfrac{\partial p_2}{\partial y} = \dfrac{1}{n}\dfrac{\partial n}{\partial y}
       $\\
    $\dfrac{\mathrm{d}x}{\mathrm{d}t} = \dfrac{p_{1}}{n^{2}}$& $\dfrac{\mathrm{d}x}{\mathrm{d}t} = \dfrac{p_{1}}{n^{2}}$\\
    $\dfrac{\mathrm{d}y}{\mathrm{d}t} = \dfrac{p_{2}}{n^{2}}$&$\dfrac{\mathrm{d}y}{\mathrm{d}t} = \dfrac{p_{2}}{n^{2}}$\\
    $\dfrac{\mathrm{d}p_{1}}{\mathrm{d}t} = \dfrac{1}{n}\dfrac{\partial n}{\partial x}$& $\dfrac{\mathrm{d}x}{\mathrm{d}t} = \dfrac{1}{n}\dfrac{\partial n}{\partial y}$
  \end{tabular}
\end{center}}

So the ODE system of four equations with four functions: $x(t)$,
$y(t)$, $p_{1}(t)$, $p_{2}(t)$, is derived:
\[
  \left\{
  \begin{aligned}
    &\dfrac{\mathrm{d}x}{\mathrm{d}t} = \dfrac{p_{1}}{n^{2}},\\
    &\dfrac{\mathrm{d}y}{\mathrm{d}t} = \dfrac{p_{2}}{n^{2}},\\
    &\dfrac{\mathrm{d}p_{1}}{\mathrm{d}t} = \dfrac{1}{n}\dfrac{\partial n}{\partial x},\\
    &\dfrac{\mathrm{d}p_{2}}{\mathrm{d}t} = \dfrac{1}{n}\dfrac{\partial n}{\partial y}.
  \end{aligned}
\right.
\]
The initial conditions:
\[
  \begin{aligned}
    &\left. x(t)\right|_{t=0} = x_{0},\\
    &\left. y(t)\right|_{t=0} = y_{0},\\
    &\left. p_{1}(t)\right|_{t=0} = c_{1}n(x_{0},y_{0}),\\
    &\left. p_{2}(t)\right|_{t=0} = c_{2}n(x_{0},y_{0}).
  \end{aligned}
\]

Constants $c_1$ and $c_2$ are bonded by following relation
$c_1^2 + c_2^2 = 1$. These constants may be presented as
$c_1 = \cos(\alpha)$ and $c_2 = \sin(\alpha)$. The initial conditions
give a mathematical description of the source of the rays. For
example, to model a point source, we need to fix the initial
coordinates $x_0$, $y_0$ and change the angle $\alpha$, which will set
the angle of the beam exit from the source-point. To simulate the
radiating surface, on the contrary, it is necessary to fix the angle
$\alpha$ and change the coordinates $x_0$ and $y_0$.

Let us to find the relation between the parameter $t$ and the function
$u(x,y)$. Since:
\[
  \frac{\mathrm{d}u}{\mathrm{d}t} = \frac{\partial u}{\partial
    x}\frac{\mathrm{d}x}{\mathrm{d}t} + \frac{\partial u}{\partial
    y}\frac{\mathrm{d}y}{\mathrm{d}t},
\]
and
\[
  \left(\frac{\partial u}{\partial x}, \frac{\partial u}{\partial
      y}\right) = \nabla u = \mathbf{p},
\]
when
\[
  \frac{\mathrm{d}u}{\mathrm{d}t} = \nabla u
  \dfrac{\mathrm{d}\mathbf{x}}{\mathrm{d}t} = \left(\mathbf{p},
    \dfrac{\mathrm{d}\mathbf{x}}{\mathrm{d}t}\right) =
  p_{1}\dfrac{\mathrm{d}x}{\mathrm{d}t} +
  p_{2}\dfrac{\mathrm{d}y}{\mathrm{d}t} = \dfrac{p_{1}p_{1}}{n^{2}} +
  \dfrac{p_{2}p_{2}}{n^{2}} = \dfrac{|\mathbf{p}|^2}{n^2}.
\]

Due to the fact that $|\mathbf{p}|^{2} = n^{2}(x,y)$, we obtain:
\[
  \frac{\mathrm{d}u}{\mathrm{d}t} = \dfrac{|\mathbf{p}|^2}{n^2} =
  \dfrac{n^2}{n^2} =
  1\;\;\Rightarrow\;\;\dfrac{\mathrm{d}u}{\mathrm{d}t} = 1.
\]

The solution of the equation $u_{t} = 1$ is the function
$u (x,y) = t + \mathrm{const}$ which implies that the parameter $t$
has a physical meaning of the signal propagation time from the point
$(x_{0}, y_{0})$ to the point $(x,y)$

In polar coordinates, the eikonal equation has the following form::
\[
\left(\dfrac{\partial u(r, \varphi)}{\partial r}\right)^2 + \dfrac{1}{r^2}\left(\dfrac{\partial u(r, \varphi)}{\partial \varphi}\right)^2 = n^2(r),
\]
and the corresponding system of ODEs will have the form:
\[
  \left\{
  \begin{aligned}
    & \dfrac{\mathrm{d}r}{\mathrm{d}t} = p_{r},\\
    & \dfrac{\mathrm{d}\varphi}{\mathrm{d}t} = \dfrac{p_{\varphi}}{r},\\
    & \dfrac{\mathrm{d}p_{r}}{\mathrm{d}t} = n\dfrac{\partial n}{\partial r} + \dfrac{p^2_{\varphi}}{r},\\
    & \dfrac{\mathrm{d}p_{\varphi}}{\mathrm{d}t} = -\dfrac{p_{\varphi}p_{r}}{r}.
  \end{aligned}
  \right.
\]
The initial conditions:
\[
  \begin{aligned}
    &\left. r(t)\right|_{t=0} = r_{0},\\
    &\left. \varphi(t)\right|_{t=0} = \varphi_{0},\\
    &\left. p_{r}(t)\right|_{t=0} = c_{1}n(r_{0}),\\
    &\left. p_{\varphi}(t)\right|_{t=0} = c_{2}n(r_{0}).
  \end{aligned}
\]

\section{Numerical simulation of Luneburg and Maxwell lenses}

Let's consider the examples of lenses~\cite{kulyabov:2017:sfm:geometrization_maxwell,kulyabov:2018:sfm:lens-calculations}.

\subsection{Luneburg lens}

Luneburg lens~\cite{luneburg:1964,morgan:1958:luneberg_lens,lock:2008:luneburg_ray,lock:2008:luneburg_wave} 
is a spherical lens of radius $R$ with the center at
point $(X_0, Y_0)$ (consider the projection on the plane Oxy) with a
refractive index of the following form
\[
  n(x,y) = 
  \left\{
  \begin{aligned}
    &n_{0} \sqrt{2-\Big(\dfrac{r}{R}\Big)^2},\quad r \leqslant R,\\
    &n_{0},\quad r > R,
  \end{aligned}
  \right.
\]
where $r(x,y) = \sqrt{(x - X_0)^2 + (y - Y_0)^2}$ is the distance from
the center of the lens to an arbitrary point in the $(x,y)$ plane. The
formula implies that the coefficient $n$ continuously varies from
$n_0\sqrt{2}$ to $n_0$ starting from the center of the lens and ending
with its boundary. The refractive index of the medium outside the lens
is constant and is equal to $n_0$. Usually $n_0$ is equal to $1$.

To solve the eikonal equation by the method of characteristics it is
necessary to find partial derivatives of the function $n(x,y)$. For
the case of Luneburg lens the partial derivatives are:
\[
  \dfrac{\partial n(x,y)}{\partial x} = -\dfrac{n_0^2 (x-X_0)}{R^2
    n(x,y)},\quad \dfrac{\partial n(x,y)}{\partial y} = -\dfrac{n_0^2
    (y-Y_0)}{R^2 n(x,y)},\; r\leqslant R.
\]
Outside the lens region derivatives are equal to $0$.

\subsection{Maxwell fish eye lens}

Maxwell fish eye lens~\cite{maxwell:1854:fish-eye} is also a
spherical lens of radius $R$ with the center at point $(X_0, Y_0)$
(consider the projection on the plane Oxy) with a refractive index of
the following form:
\[
  n(x,y) = 
  \left\{
  \begin{aligned}
    &\dfrac{n_{0}}{1+\Big(\dfrac{r}{R}\Big)^2},\quad
    r \leqslant R,\\
    &n_{0},\quad r > R.
  \end{aligned}
  \right.
\]

To solve the eikonal equation by the method of characteristics it is
necessary to find partial derivatives of the function $n(x,y)$. For
the case of Maxwell lens partial derivatives have the form:
\[
  \dfrac{\partial n(x,y)}{\partial x} = -\dfrac{2n^2(x,y) (x-X_0)}{n_0
    R^2},\quad \dfrac{\partial n(x,y)}{\partial y} = -\dfrac{2n^2(x,y)
    (y-Y_0)}{n_0 R^2},\quad r\leqslant R.
\]

\subsection{Description of the numerical modeling}

Julia programming language~\cite{joshi:book:learning-julia} is used to
simulate the trajectories of rays through the Maxwell and Luneburg
lenses. We use classical Runge--Kutta methods with constant step to
solve the ODE system.

We carry on numerical modeling for lenses with a radius $R = 1$, the
refractive index of the external medium $n_0 = 1$, the center of the
lens was placed in the point $(X_0, Y_0) = (2, 0)$, the boundary
region was set as the rectangle $x_{\min} = 0$, $x_{\max} = 5$,
$y_{\min}=-1.5$ and $y_{\max} = 1.5$. The point source was placed on
the lens boundary at $(x_0, y_0) = (0, 0)$. $50$ values of the
$\alpha$ parameter have been taken from the interval
$[-\pi/2 + \pi/100, \pi/2 - \pi/100]$, which allowed to simulate rays
trajectories from a point source within an angle slightly smaller than
$180^{\circ}$. The $t$ parameter was changed within the $[0, 5]$
interval.

\begin{figure}
  \begin{minipage}{0.45\linewidth}
    \centering
    \includegraphics[width=\linewidth]{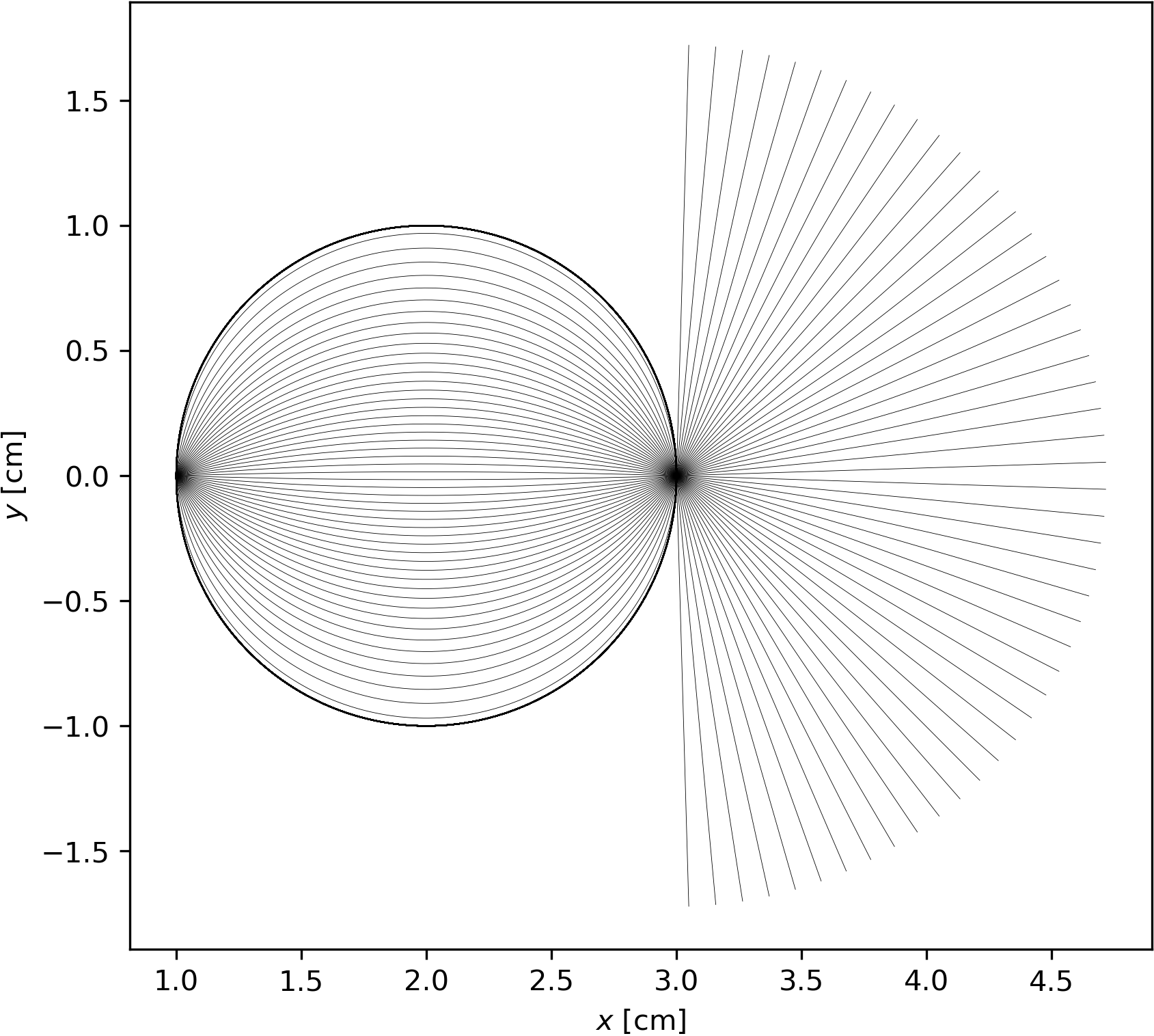}
    \caption{The trajectories of the rays in case of Maxwell's lens for a point source and $n_0 = 1$}
    \label{fig:maxwell_rays}
  \end{minipage}
  \hfill
  \begin{minipage}{0.45\linewidth}
    \centering
    \includegraphics[width=\linewidth]{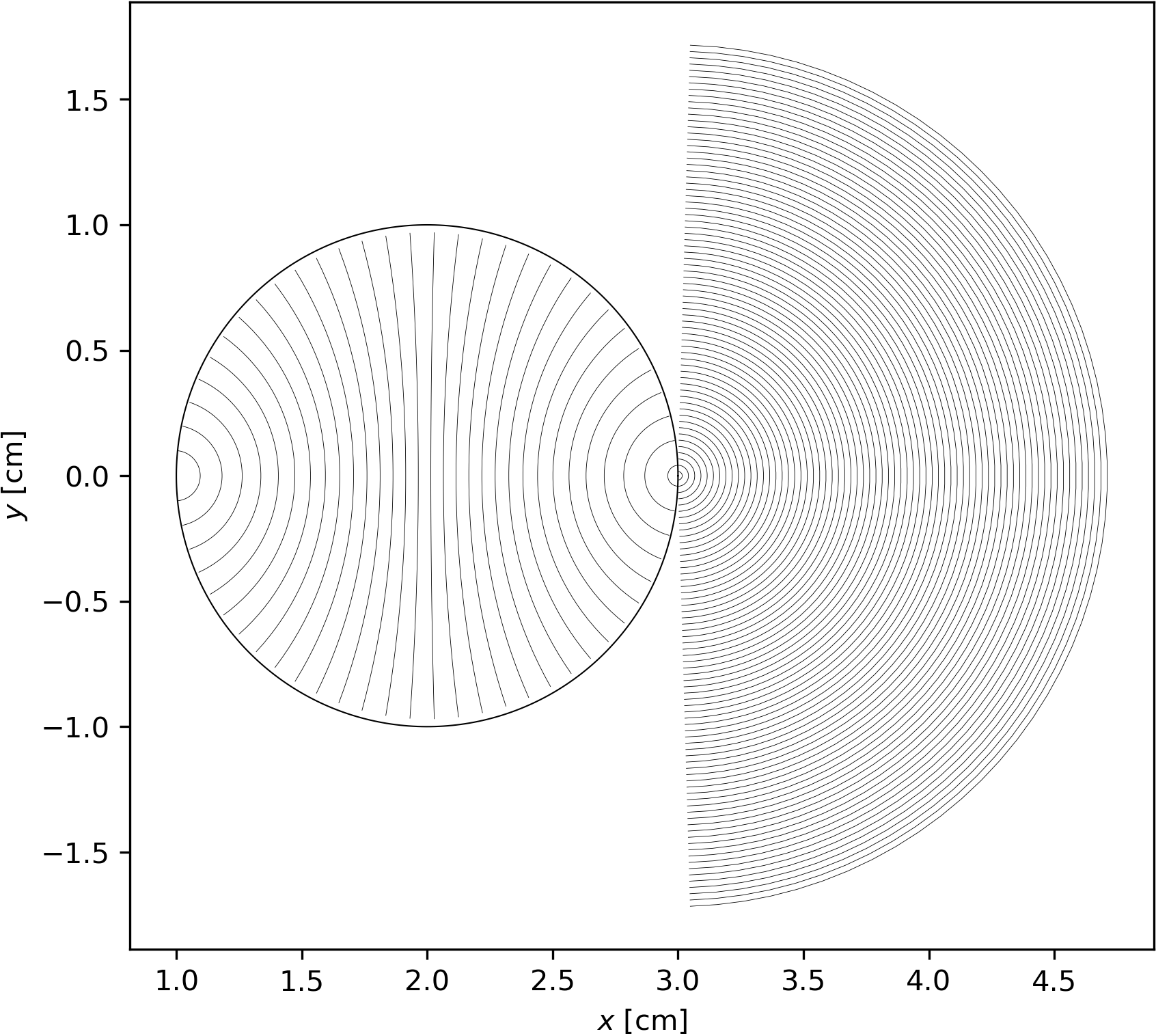}
    \caption{The wavefronts for in case of Maxwell's lens for a point source and $n_0 = 1$}
    \label{fig:maxwell_fronts}
  \end{minipage}
\end{figure}

\begin{figure}
  \centering
  \includegraphics[width=0.8\linewidth]{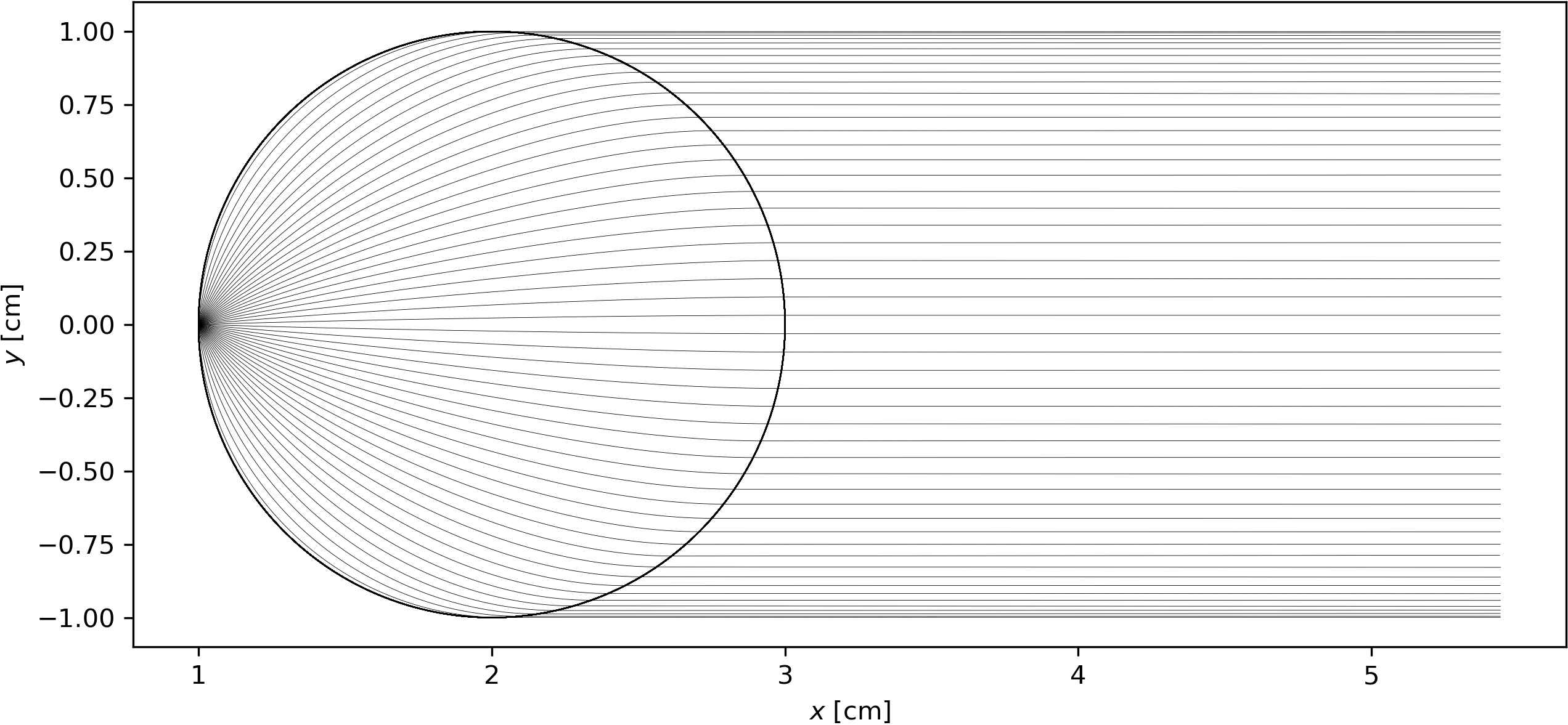}
  \caption{The trajectories of the rays in case of Luneburg lens for a point source and $n_0 = 1$.}
  \label{fig:luneburg_rays}
\end{figure}

\begin{figure}
  \centering
  \includegraphics[width=0.8\linewidth]{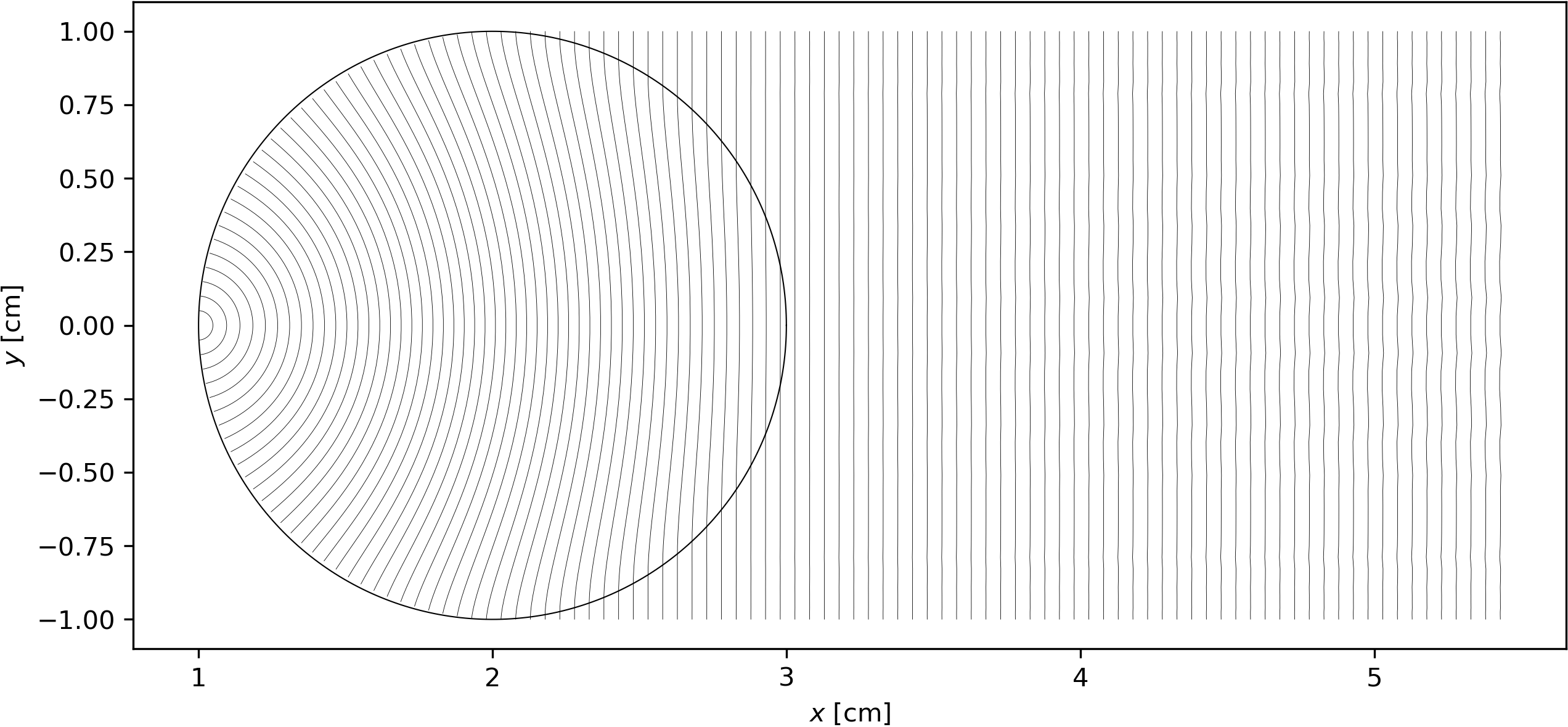}
  \caption{The wavefronts in case of Luneburg lens for a point source and $n_0 = 1$.}
  \label{fig:luneburg_fronts}
\end{figure}

Each $\alpha$ parameter value sets new initial conditions for the ODE
system. The process of numerical simulation consists in multiple
solution of this system for different initial conditions. The
numerical solution of the ODE system for a particular initial
condition gives us a set of points $(x_{I},y_{I})$, $I=1,\ldots,N$
approximating the trajectories of a particular beam. After performing
calculations for all the selected initial conditions, we obtain a set
of rays. To visualize the rays, it is enough to depict each of the
obtained numerical solutions. The result of the simulation can be seen
in the Fig.~\ref{fig:maxwell_rays} and Fig.~\ref{fig:luneburg_rays} (the
trajectories of the rays) and Fig.~\ref{fig:maxwell_fronts} and
Fig.~\ref{fig:luneburg_fronts} (the wavefronts).

  
To visualize the wave fronts with the resulting numerical data it is
necessary to carry out additional recalculations. From each numerical
solution, we must select points $(x_{I},y_{I})$ that correspond to a
specific point in time $t_{I}$.

The use of a numerical method with a fixed step gives an advantage,
since each numerical solution will be obtained for the same uniform
grid ${t_0 < t_1 < \ldots < t_i < \ldots < t_n}$.

\section{Conclusion}
\label{sec:conclusion}

The paper presents the description of the numerical solution of the
eikonal equation for the case of Luneburg and Maxwell lenses. The
results are visualized as trajectories of rays passing through lenses
and as fronts of electromagnetic waves.

\begin{acknowledgments}

The publication has been prepared with the support of the ``RUDN University Program 5-100''
and funded by Russian Foundation for Basic Research (RFBR) according to the research project
No~19-01-00645.

\end{acknowledgments}

 \bibliographystyle{elsarticle-num}

\bibliography{bib/eikonal-sfm/cite}

\end{document}